\documentclass[journal,comsoc]{IEEEtran}
\usepackage[T1]{fontenc}

\usepackage{cite}

\usepackage[pdftex]{graphicx}
\graphicspath{{Figures/}}

\usepackage{amsmath}
\usepackage[cmintegrals]{newtxmath}
\usepackage{enumitem}
\setlist[description]{wide=\parindent}
\usepackage{algorithmic}
\usepackage{algorithm}

\usepackage{array}

\usepackage[caption=false,font=normalsize,labelfont=sf,textfont=sf]{subfig}

\usepackage[dvipsnames]{xcolor}
\begin{document}
%
\title{End-to-end Learning of a Constellation Shape Robust to Channel Condition Uncertainties}
%
%
%

\author{Ognjen~Jovanovic,~\IEEEmembership{Graduate Student Member,~IEEE,}
        Metodi~P.~Yankov,~\IEEEmembership{Member,~IEEE,}
        Francesco~Da~Ros,~\IEEEmembership{Senior Member,~IEEE,}
        and~Darko~Zibar
\thanks{O. Jovanovic, M. P. Yankov, F. Da Ros, and D. Zibar are with the Department
of Photonic Engineering, Technical University of Denmark, 2800 Kgs. Lyngby,
Denmark, e-mail: ognjo@fotonik.dtu.dk}
}

%
%

\markboth{ECOC extention}%
{Preprint}
%


\IEEEspecialpapernotice{(Invited Paper)}

\maketitle

\begin{abstract}
Vendor interoperability is one of the desired future characteristics of optical networks. This means that the transmission system needs to support a variety of hardware with different components, leading to system uncertainties throughout the network. For example, uncertainties in signal-to-noise ratio and laser linewidth can negatively affect the quality of transmission within an optical network due to e.g. mis-parametrization of the transceiver signal processing algorithms. In this paper, we propose to geometrically optimize a constellation shape that is robust to uncertainties in the channel conditions by utilizing end-to-end learning. In the optimization step, the channel model includes additive noise and residual phase noise. In the testing step, the channel model consists of laser phase noise, additive noise and blind phase search as the carrier phase recovery algorithm. Two noise models are considered for the additive noise: white Gaussian noise and nonlinear interference noise model for fiber nonlinearities. The latter models the behavior of an optical fiber channel more accurately because it considers the nonlinear effects of the optical fiber. For this model, the uncertainty in the signal-to-noise ratio can be divided between amplifier noise figures and launch power variations. For both noise models, our results indicate that the learned constellations are more robust to the uncertainties in channel conditions compared to a standard constellation scheme such as quadrature amplitude modulation and standard geometric constellation shaping techniques.
\end{abstract}

\begin{IEEEkeywords}
Optical fiber communication, end-to-end learning, geometric constellation shaping, phase noise.
\end{IEEEkeywords}

%
\IEEEpeerreviewmaketitle

\section{Introduction}
\IEEEPARstart{O}{ptical} networks have to continuously evolve to keep up with the growth of data traffic demand. To efficiently meet the demand, the optical communication systems have to offer higher spectral efficiency. Geometric constellation shaping (GSC) may be used to optimize high-order modulation formats to improve the spectral efficiency and maximize the mutual information (MI) for the given channel. However, including all noise sources present in the entire chain of coherent optical communication system is difficult and often disregarded for such optimization. One of the noise source that should be included is the residual phase noise (RPN) which is the result of an imperfect carrier phase estimation (CPE) and compensation of the laser phase noise (PN). 

The parametrization of CPE algorithms is sensitive to channel conditions, such as the signal-to-noise ratio (SNR) and the linewidth (LW) which might be challenging to measure in practical scenarios. Furthermore, variations in the channel conditions throughout the network can occur, e.g. due to vendor interoperability and aging. Network operators are seeking vendor interoperability over their networks\cite{Filer:18,Gunkel} to reduce the cost of the infrastructure. This means that the transmission needs to support a variety of different optical network elements. Parameters characterizing these elements, such as amplifier noise figures (NF) and laser linewidth (LW) can be vendor dependent and vary throughout the network. Also, as a result of aging, the launch powers may shift over time. Due to these variations, the knowledge of the channel conditions is imperfect and might lead to a mis-parametrized digital signal processing (DSP) blocks, such as the CPE. 

In probabilistic shaping (PS), the constellation shape is by definition changed with the SNR and the target data rate \cite{Bocherer_TCOM15}. There have been some studies on the robustness of a chosen shape to SNR variation \cite{Fehenberger:16}, and also how to select the shapes to be robust to RPN \cite{Mello_JLT2018}. In general, PS requires a distribution matcher to be implemented, which leads to higher complexity of the transceiver. Instead, GCS is directly compatible with classical bit-interleaved coded modulation (BICM). Therefore, finding a robust GCS that maintains good performance under channel condition uncertainties is of the utmost importance. A possible strategy to perform GCS is by utilizing end-to-end learning.

End-to-end learning was introduced in \cite{OShea2017}, where it was shown that a communication system (or some of its processes) can be optimized for a specific channel and performance metric by utilizing a deep learning concept known as autoencoders (AEs) \cite{goodfellow2016deep}. End-to-end learning has been applied in optical communication for GCS \cite{Jones2018a,Jones2019,Li2018b,Schaedler2020,Gumus2020,Vlad_ECOC,oliari2021highcardinality} mainly focusing on the mitigation of the nonlinear effects of the optical fiber. Apart from GCS, end-to-end learning was applied in optical communication for waveform optimization for dispersive fiber\cite{Karanov2018,Karanov2019a,Karanov2021}, waveform optimization for nonlinear frequency division multiplexing\cite{Gaiarin2020,GaiarinJLT} and superchannel transmission\cite{Song:21}. In \cite{Karanov2018,Karanov2021}, they have demonstrated that by varying the distance in the optimization process, the learned waveform is robust to variations in transmission distances for short-reach intensity-modulated communication systems.

Performing GCS using end-to-end learning, which usually relies on gradient-based optimization, can be difficult on a channel that includes CPE because it is usually complex and non-differentiable, e.g. the blind phase search (BPS) algorithm\cite{Pfau2009a}. It was shown that this optimization is possible by using a gradient-free optimization method \cite{Ognjen_JLT}. Recently, a differentiable version of the BPS was proposed in \cite{rode2021geometric}. However, a more typical approach, used in previous works on GCS that do not use end-to-end learning \cite{Li,Pfau,Krishnan,Kayhan,Dzieciol}, is artificially modeling the RPN. These previous works assume ideal knowledge of the channel conditions which does not reflect the true RPN after the mis-parametrized CPE.

This paper is an extension of \cite{Ognjen_ECOC}, which shows how to learn a constellation robust to signal-to-noise ratio (SNR) and laser linewidth (LW) uncertainties by utilizing end-to-end learning. In this paper, the results from \cite{Ognjen_ECOC} are discussed and the analysis is extended also to an optical communication channel modeled with the nonlinear interference noise (NLIN) model\cite{Dar2014}. The launch power and the amplifier noise figure (NF) uncertainties are considered as the main cause of the SNR uncertainty within optical networks\cite{Seve:18}. Therefore, a constellation shape that is robust to variations in the launch power, the amplifier NF and the laser LW is learned in this case. This constellation is learned by varying the launch power, the amplifier NF and the RPN severity in a simple differentiable channel during the training step. Afterwards, the constellation is tested on a more realistic channel, that includes laser PN, NLIN and BPS, which better reflects the true RPN. 

The remainder of the paper is organized as follows. Mutual information (MI) is used as the performance metric and the basic principles of estimating it are described in Section \ref{MI}. A detailed description of the training and testing setups is provided in Section \ref{Meth}. Section \ref{channel_models} describes the used channel models and different optimization scenarios for each of them. Section \ref{results} provides the results on the mutual information obtained by different constellations in the testing setup. The conclusions are summarized in Section \ref{Conclusion}.


\section{Performance metric}\label{MI}

Let $\mathcal{X}$ be a set of complex constellation points (symbols) with cardinality $|\mathcal{X}|=M=2^m$, where $m$ is the number of bits carried by a symbol. Consider $X$ and $Y$ to be the input and the output sequence of a communication channel, respectively and that their relation is governed by the channel transition probability density $p_{Y|X}(y|x)$. The symbol sequence $X$ is sampled from $\mathcal{X}$ with a uniform probability mass function $P_{X}(x)=\frac{1}{M}$ and has entropy $H(X)= -\sum_{x \in \mathcal{X}}P_{X}(x)\log_2(P_{X}(x))=\log_2(M)=m$. The sequence $Y \in \mathbb{C}$ and has a probability distribution $p_{Y}(y)$, where $\mathbb{C}$ denotes the set of complex numbers. The conditional entropy of $X$ given $Y$ is $H_p(X|Y)= \mathbb{E}_{p(x,y)}[p_{X|Y}(x|y)]$. The expectation $\mathbb{E}_{p(x,y)}$ should be taken over the true joint probability density function of $p_{X,Y}(x,y)$. The entropy $H(X)$ and conditional entropy $H_p(X|Y)$ can be used to calculated the amount of information $Y$ contains about $X$ in bits per symbol 
\begin{equation}
\begin{split}
    I(X;Y) & = H(X)-H_p(X|Y) = m - H_p(X|Y) \\
    & = \sum_{x \in \mathcal{X}}P_{X}(x) \int_{\mathbb{C}} p_{Y|X}(y|x)\log_2 \frac{p_{Y|X}(y|x)}{p_{Y}(y)}dy \text{,} 
\end{split}
\label{eq:MI}
\end{equation}
known as MI $I(X;Y)$. Even though $P_{X}(x)$ is a uniform probability mass function, it is included to the equation because the method could be extended to other probability mass functions, e.g. optimized for probabilistic shaping.

In order to evaluate Eq. (\ref{eq:MI}), the transition probability $p_{Y|X}(y|x)$ must be known and this is usually not the case. A typical approach when $p_{Y|X}(y|x)$ is unknown, is to bound Eq. (\ref{eq:MI}). The mismatched decoding approach can be used to obtain a lower bound on the MI. It assumes the transition probability $q_{Y|X}(y|x)$ of an auxiliary channel instead of the true $p_{Y|X}(y|x)$ \cite{Lowerbound}. Then a lower bound on the MI, also known as the achievable information rate (AIR), is formulated as 
\begin{equation}
I(X;Y) \geq H(X)-\hat{H}_q(X|Y) = m - \hat{H}_q(X|Y)\text{,}
\label{eq:MI_lower}
\end{equation}
where $\hat{H}_q(X|Y)=\mathbb{E}_{p(x,y)}[q_{X|Y}(x|y)]$ is the upper bound of the true conditional entropy $H_p(X|Y)$. The inequality in Eq. (\ref{eq:MI_lower}) turns to equality when $q_{Y|X}(y|x)=p_{Y|X}(y|x)$.

\section{Methodology}\label{Meth}

\subsection{Geometric constellation shaping with autoencoders}

\begin{figure}[!t]
\centering
\subfloat[]{\includegraphics[width=\columnwidth]{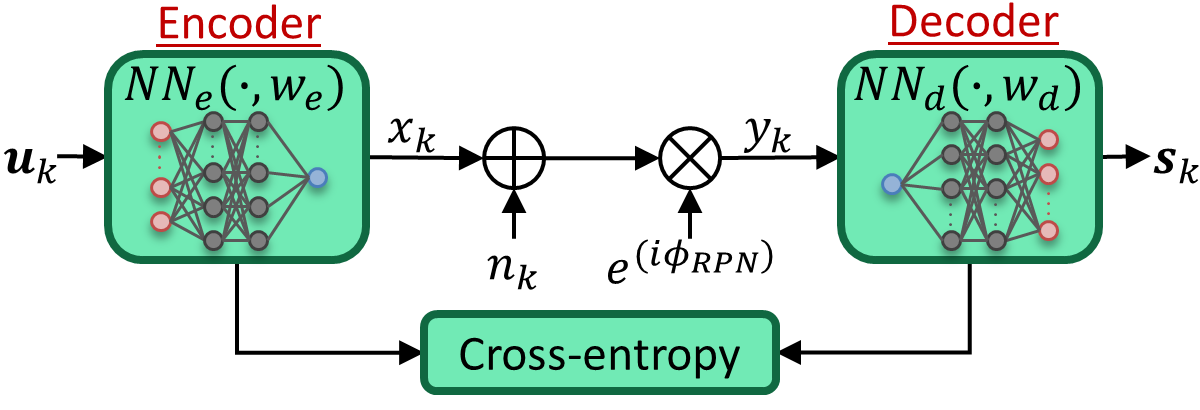}}
\hfil
\subfloat[]{\includegraphics[width=\columnwidth]{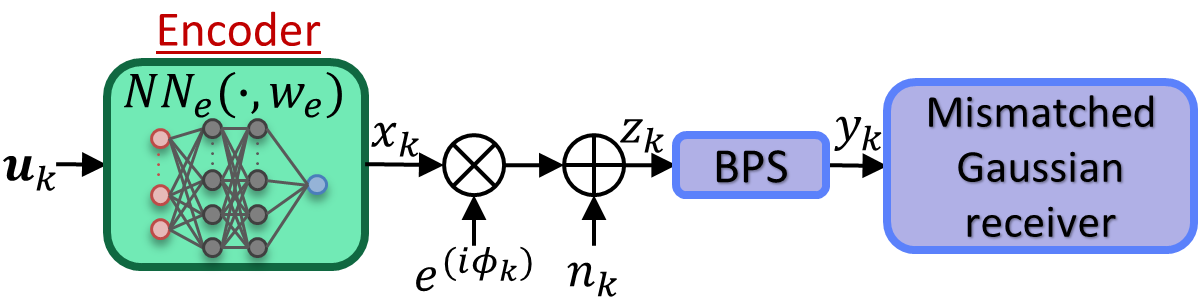}}
\caption{The illustration of the: (a) training setup for geometrical constellation shaping which is an autoencoder with an embedded channel model. The embedded channel model consists of additive noise and residual phase noise; (b) testing setup of the learned constellation. The setup consists of the trained encoder (constellation under test), phase noise and additive noise channel, BPS as the phase recovery algorithm, and mismatched Gaussian receiver to estimate the mutual information.}
\label{fig:Train}
\end{figure}

Geometric constellation shaping may be used to optimize the position of constellation points in a high-order modulation formats to improve the spectral efficiency and maximize the MI $I(X;Y)$. The considered training setup employing an AE for GCS is shown in Fig. \ref{fig:Train} (a). An AE consists of two neural networks (NNs), an encoder and a decoder with a channel model in between. The encoder and the decoder are represented by feed-forward neural networks $NN_{e}(\cdot,\mathbf{w}_{e})$ and $NN_{d}(\cdot,\mathbf{w}_{d})$, parameterized with trainable weights (including biases) $\mathbf{w}_{e}$ and $\mathbf{w}_{d}$, respectively. The overall goal is to find the weight set, $\mathbf{w}=\{\mathbf{w}_e,\mathbf{w}_d\}$, that would minimize the cross-entropy between the input and output of the AE for the considered channel. The encoder optimizes the position of the constellation points, whereas the decoder learns the decision boundaries of the distorted symbols.

The input to the encoder is a one-hot encoded vector $\mathbf{u}_k \in \mathbb{U} = \{ \mathbf{e}_i | i=1,\dots, M\}$ which is mapped to a normalized complex constellation point $x_k=NN_e(\mathbf{u}_k,\mathbf{w}_{e})$, where $k$ represents the $k$-th sample and $\mathbf{e}_i$ is an all zero vector with a one at position $i$. The output of the network is two-dimensional, representing the real and imaginary part of the complex constellation symbol $x_k$. During training, the complex symbol $x_k$ is transmitted over a one sample per symbol channel model with symbol rate $R_s$, which consists of complex additive noise $n_k$ and multiplicative RPN $ \phi_{RPN}$, resulting into the impaired symbol $y_k = (x_k+n_k)e^{i \phi_{RPN}}$. The additive noise $n_k$ and the RPN $\phi_{RPN}$ are modeled as zero-mean Gaussian distributions with variances $\sigma_n^2$ and $\sigma_{RPN}^2$, respectively. It should be emphasized that the additive noise $n_k$ is complex valued and circular symmetric, whereas the RPN $\phi_{RPN}$ is real valued. The real and the imaginary part of the impaired symbol $y_k$ are inputs to the decoder, which outputs a vector of posterior probabilities $\mathbf{s}_k=NN_d(y_k,\mathbf{w}_{d}) \in [0,1]^M$ using a softmax output layer. The optimization of the AE weight set $\mathbf{w}$ is performed by iteratively minimizing the cross-entropy cost function over a sample set of size $N$. In each iteration, the sample set is divided into batches of size $B$ and the cross-entropy loss for each batch is calculated as
\begin{equation}
    \begin{split}
    J_{CE}(\mathbf{w}) & = \frac{1}{B} \sum_{k=1}^{B} \bigg[ -\sum_{i=1}^{M} \mathbf{u}^{(i)}_k \log \mathbf{s}^{(i)}_k \bigg]
    \end{split}
    \label{eq:Xent}
\end{equation}
where $(i)$ denotes the $i$-th element of the vector. The output of the decoder is an approximation $q_{X|Y}(x|y)$ of the true posterior distribution $p_{X|Y}(x|y)$. Therefore, the cross-entropy can used to calculate the AE-based upper bound on the conditional entropy $\hat{H}_{q}(X|Y)=\mathbb{E}_{p(x,y)}[q_{X|Y}(x|y)]$. Based on Eq. (\ref{eq:MI_lower}), this implies that minimizing the cross-entropy maximizes a lower bound on the MI. In particular, this lower bound is an AIR when using the decoder NN. Once the training has converged, the encoder weights are fixed and the testing is performed. The AE hyperparameters are shown in Table \ref{tab:Setup}.

The MI is the best performance a communication system can attain, however it reflects the achievable information rate in the cases of iterative demapping and decoding or non-binary forward error correction (FEC). In optical communication, BICM is usually used and it requires a constellation with a Gray-like labeling. Therefore, optimizing a constellation based on MI could result in penalty in the actual achievable rate. The current system could be expanded such that it includes optimization of the bit labeling as done in \cite{Jones2019,rode2021geometric} and this is left for future work.

\subsection{Testing setup for the learned constellation}

The learned constellations are tested on the setup shown on Fig. \ref{fig:Train} (b) which is more realistic than the one used for training. The encoder output $x_k$ is transmitted over a channel consisting of laser PN $\phi_k$ and additive noise $n_k$, resulting into an impaired symbol 
\begin{equation}
    z_k = x_k e^{i \phi_k} + n_k \text{.}
\end{equation}
The additive noise is distributed identically to the noise during training and the laser PN is modeled as a Wiener process
\begin{equation}
    \phi_k = \phi_{k-1} + \Delta \phi_k \text{,}
\end{equation}
where $\Delta \phi_k$ is the random phase increment sampled from a zero-mean Gaussian with variance  $\sigma^2_\phi = 2\pi\Delta\nu T_s$. The combined transmitter and receiver laser LW is denoted by $\Delta \nu$, and $T_s=1/R_s$ is the symbol period.

At the receiver, the laser PN is estimated with a blind phase search (BPS) \cite{Pfau2009a}, which is a standard phase noise compensation algorithm. The BPS is a pure feedforward phase recovery algorithm which estimates the phase by rotating the received symbol by $N_s$ test phases defined by
\begin{equation}
    \theta_j=\frac{j}{N_s}\cdot 2\pi, \quad j \in \{0,1,\dots,N_s-1 \},
\end{equation}
where $j$ represents the $j$-th test phase. Each of the rotated symbols $z_{k,j} = z_k e^{-i\theta_j}$ is fed into a minimum distance decision operator to determine the closest symbol. The distance between the decided symbol $\hat{z}_{k,j}$ and the rotated symbol $z_{k,j}$ is calculated as
\begin{equation}
    d_{k,j}=|z_{k,j}-\hat{z}_{k,j}|\text{.}
\end{equation}
In order to mitigate the effect the additive noise has on the performance of the phase recovery, the squared distances of symbols rotated by the same test phase are summed over a window of size $2W+1$
\begin{equation}
    r_{k,j}=\sum_{i=-W}^{W}d_{k-i,j}^2\text{.}
\end{equation}
Finally, the optimal test phase is chosen by the minimum sum of squared distances \cite{Pfau2009a}
\begin{equation}
    \hat{\phi}_k=\underset{\theta_j}{\mathrm{argmin}} \quad r_{k,j},
    \label{eq:hd}
\end{equation}
where $\mathrm{argmin}$ is a non--differentiable operation.
The received symbol is rotated by the chosen test phase to output the phase compensated sample
\begin{equation}
    y_k=z_k e^{-i\hat{\phi}_k}.
\end{equation}
The BPS algorithm is non-differentiable due to its hard-decision directed nature given by Eq. (\ref{eq:hd}). Therefore, the gradient of the BPS algorithm cannot be computed, making it difficult to use for training that relies on gradient-based optimization. Instead, the RPN as described above is adopted during training.

\begin{table}[!t]
\renewcommand{\arraystretch}{1.3}
\caption{Parameters of the encoder and decoder neural network}
\label{tab:Setup}
\centering
\begin{tabular}{|c||c|c|}
\hline
 & Encoder NN & Decoder NN\\
\hline
\# of input nodes & $M$ & 2\\
\hline
\# of hidden layers & 0 & 1\\
\hline
\# of nodes per hidden layer & 0 & $M/2$\\
\hline
\# of output nodes & 2 & $M$\\
\hline
Bias & No & Yes\\
\hline
Hidden layer activation function & None & Leaky Relu\\
\hline
Output layer activation function & Linear & Softmax\\
\hline
\end{tabular}
\end{table}

For optical communication, it is a common approach to use a mismatched Gaussian receiver \cite{Lapidoth1994,Metodi:16} to estimate the MI between the channel input and output. The mismatched Gaussian receiver assumes the transition probability $q_{Y|X}(y|x)$ in Eq. (\ref{eq:MI_lower}) is of an auxiliary Gaussian channel
\begin{equation}
   q_{Y|X}(y|x) = \frac{1}{\sqrt{2 \pi \sigma_{G}^2}}\exp{\left(-\frac{|y-x|^2}{2 \sigma_{G}^2}\right)} \text{,}
\end{equation}
where $\sigma_{G}^2$ is the estimated noise variance of the auxiliary channel. Applying the Bayes' theorem, the posterior distributions are
\begin{equation}
    q_{X|Y}(x|y) = \frac{p_X(x)q_{Y|X}(y|x)}{\sum_{x^i \in \mathcal{X}} p_X(x=x^i)q_{Y|X}(y|x=x^i)} \text{.}
    \label{eq:qxy}
\end{equation}
The combined distortion of the the additive noise and the RPN is assumed to be purely Gaussian and parametrized by the noise variance $\sigma_{G}^2$ estimated from the received sequence as $\sigma_{G}^2=\mathbb{E}[|y-x|^2]$. Then, the Monte Carlo approach can be used to evaluate Eq. (\ref{eq:qxy}). The auxiliary function $q_{X|Y}(x|y)$ is an approximation to $p_{X|Y}(x|y)$ in two ways: 1) it is modeled using a decoder NN or a Gaussian receiver; 2) it is memoryless. Both of these approximations lead to an upper bound on the conditional entropy and a lower bound on the MI.

It should be mentioned that in the training scenario, the RPN is added after the additive noise because it occurs as a result of an imperfect compensation of the PN at the receiver. In the testing scenario, ideally, there should be two PN sources, the transmitter and the receiver laser. Since the two processes are independent of each other, independent of the additive noise, and they do not alter the circular Gaussian distribution of the additive noise, they can be combined into a single process with a variance that is the sum of the two variances. The resulting PN may be added on either side of the additive noise without changing the statistics of the channel model.

\section{Additive noise models} \label{channel_models}
In this paper, two additive noise models are used, additive white Gaussian noise (AWGN) and nonlinear interference noise (NLIN) model for fiber communication. 
\subsection{Additive white Gaussian noise} \label{ss:AWGN}

The noise variance in the case of the AWGN is determined by the signal-to-noise ratio (SNR): $\sigma_n^2=\frac{1}{SNR}$. The training scenarios for the AWGN are \cite{Ognjen_ECOC}: 
\begin{description} [itemsep=4pt]
    \item[AWGN 1)] Constellations trained on SNR and RPN variance $\sigma^2_{RPN}$ pairs which are fixed, similar to what was done in \cite{Dzieciol}. The SNR values were chosen from a set SNR~$\in\{15,16,...,20\}$~dB and the RPN variance is taken from a coarsely chosen set $\sigma^2_{RPN} \in \{0.001,0.005,0.01,0.02,0.05\}$.
    Therefore, a constellation is learned for each SNR and $\sigma^2_{RPN}$ combination. The best performing constellations with regards to MI for known SNR and laser LW pairs should be found this way.
    
    \item[AWGN 2)] A constellation trained on a fixed SNR and varying RPN, resulting in a constellation robust to laser LW uncertainties for a fixed SNR. The SNR$=17$~dB and the RPN variance is sampled each training batch from a log-uniform distribution in the range of $\sigma^2_{RPN} \in [0.005, 0.02]$.
    
    \item[AWGN 3)] A constellation trained on varying SNR and RPN, resulting in a constellation robust to both SNR and LW uncertainties. Each training batch, the SNR is sampled from a uniform distribution SNR $\in [15, 20]$~dB and the RPN variance is sampled from a log-uniform distribution $\sigma^2_{RPN} \in [0.005, 0.05]$.
\end{description}

\begin{table}[!t]
\renewcommand{\arraystretch}{1.5}
\newcolumntype{C}[1]{>{\centering\let\newline\\\arraybackslash\hspace{0pt}}m{#1}}
\caption{Channel parameters}
\label{tab:channel}
\centering
\begin{tabular}{| C{0.45\columnwidth} | C{0.45\columnwidth}|}
\hline
 Symbol rate ($R_s$) & 32 GHz\\
\hline
Carrier frequency ($F_c$) & 193.41 THz\\
\hline
\# of channels & 5\\
\hline
Channel spacing & 50 GHz\\
\hline
\# of polarizations ($N_{pol}$) & 2\\
\hline
\# of spans ($N_{sp}$) & 10\\
\hline
Span length ($L$) & 100 km\\
\hline
Attenuation ($\alpha$) & 0.2 dB/km\\
\hline
Amplifier gain ($G$) & $L\cdot \alpha$\\
\hline
Nonlinear coefficient & $1.3(\text{W km})^{-1}$\\
\hline
Dispersion parameter & 16.464 ps/(nm km)\\
\hline
\end{tabular}
\end{table}

\subsection{Nonlinear interference noise} \label{ss:NLIN}
Constellations optimized with regards to AWGN can be sub-optimal for a nonlinear channel such as the optical fiber. An optimal constellation for the optical channel should be jointly robust to amplification noise and signal dependent nonlinear interference \cite{Metodi:16}. The NLIN model \cite{Dar2014} for fiber communication takes into account the nonlinear interference dependent on the launch power per channel and the moments of the constellation. The NLIN model assumes that the nonlinear effects degrading the transmitted signal can be modeled as additive Gaussian noise for which the variance is determined by the parameters of the fiber communication channel. Based on this model, the channel impairments depend on the amplified spontaneous emission (ASE) noise governed by the amplifier noise figure $F_n$, the average launch power per channel $P_{in}$ (in the rest of the paper referred to as launch power) and the high order moments of the constellation,
\begin{equation}
    \mu_4 = \frac{\mathbb{E}[|X|^4]}{(\mathbb{E}[|X|^2])^2} \quad \text{and} \quad
    \mu_6 = \frac{\mathbb{E}[|X|^6]}{(\mathbb{E}[|X|^2])^3}\text{.}
\end{equation}
In the case of the NLIN model, the noise variance is defined as
\begin{equation}
    \sigma_n^2 = \sigma_{ASE}^2(F_n) + \sigma_{NLIN}^2(P_{in},\mu_4,\mu_6) \text{,}
    \label{eq:noise}
\end{equation}
where $\sigma_{ASE}^2(F_n)$ is the variance of the ASE noise and $\sigma_{NLIN}^2(P_{in},\mu_4,\mu_6)$ is the variance of the nonlinear interference. Other parameters of the optical channel that contribute to the noise variances are not included in Eq. (\ref{eq:noise}) because they are fixed. These parameters are provided in Table \ref{tab:channel}.

In optical networks, the SNR uncertainty is essentially due to uncertainties in launch power $P_{in}$, that normally has values in the interval $[-2,2]$ dBm, and amplifier noise figure $F_n$ which normally has values in the interval $[5,7]$ dB \cite{Seve:18}. It should be mentioned that for this channel model the encoder output $x_k$ is rescaled by the launch power $P_{in}$ such that $x_k=\sqrt{P_{in}} \cdot NN_e(\mathbf{u}_k,\mathbf{w}_{e})$. The training scenarios for NLIN are:

\begin{figure}[!t]
\centering
\includegraphics[width=\columnwidth]{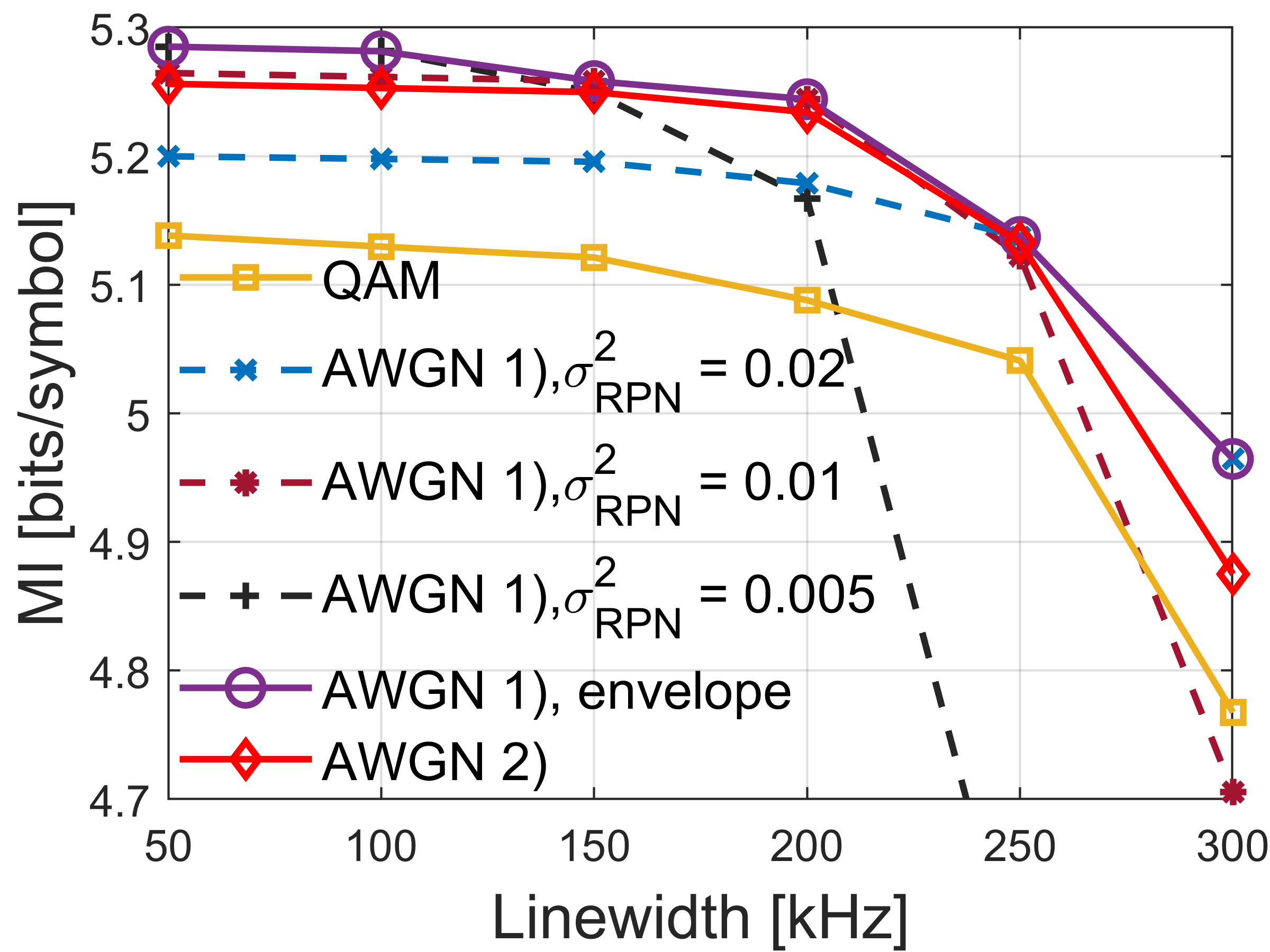}
\caption{Performance in MI with respect to LW at SNR~$=17$~dB for QAM, AWGN 1) and AWGN 2), where AWGN 1) and AWGN 2) represent scenarios 1) and 2) for the AWGN model, respectively. The AWGN 1) is the scenario where both the SNR and the RPN are fixed values, whereas the AWGN 2) is the scenario where the SNR is fixed and the RPN value varies during training.}
\label{fig:LW_robust_MI}
\label{fig:AWGN}
\end{figure}

\begin{figure*}[!t]
\centering
\includegraphics[width=\linewidth]{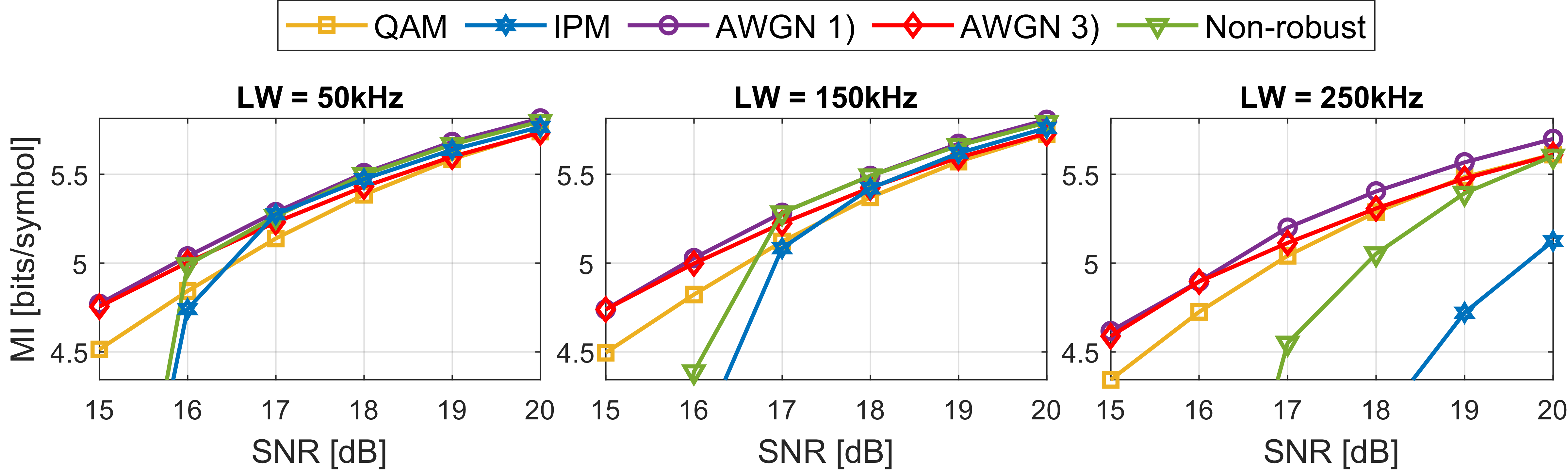}
\caption{Performance in MI with respect to SNR for different LW values for QAM, IPM, AWGN 1) and AWGN 3), where AWGN 1) and AWGN 3) represent scenarios 1) and 3) for the AWGN model, respectively. The AWGN 1) is the scenario where both SNR and RPN are fixed values, whereas the AWGN 3) is the scenario where both the SNR and the RPN values vary during training. The constellation denoted as non-robust is optimized for the pair SNR~$=18$~dB and $\Delta \nu =150$~kHz.}
\label{fig:LW_SNR_robust_MI}
\end{figure*}

\begin{description}[itemsep=4pt]
    \item[NLIN 1)] Constellations trained on a fixed noise figure $F_n$, launch power $P_{in}$ and RPN variance $\sigma^2_{RPN}$ triplets. The noise figure values are chosen from a set $F_n \in \{5,6,7\}$~dB, the launch power values are chosen from a set $P_{in} \in \{-2,-1.5,...,2\}$~dBm and RPN variance is taken from a coarsely chosen set $\sigma^2_{RPN} \in \{0.001,0.005,0.01,0.02,0.05\}$. Therefore, a constellation is learned for each $F_n$, $P_{in}$ and $\sigma^2_{RPN}$ combination. Similarly to scenario 1) for AWGN model, the goal is to find a best performing constellation with regards to MI for a given noise figure $F_n$, launch power $P_{in}$ and laser LW triplet.
    
    \item[NLIN 2)] Constellations trained on a fixed noise figure $F_n$, varying $P_{in}$ and RPN variance $\sigma^2_{RPN}$, resulting in a constellation for each $F_n$ value robust to $P_{in}$ and laser LW uncertainties. The noise figure values are chosen from a set $F_n \in \{5,6,7\}$~dB, for each training batch the launch power is sampled from a continuous uniform distribution  $P_{in} \in [-2, 2]$~dBm and the RPN variance is sampled from a log-uniform distribution $\sigma^2_{RPN} \in [0.005, 0.05]$. In this scenario three constellation are learned, one for each noise figure value.
    
    \item[NLIN 3)] A constellation trained on varying NF, $P_{in}$ and RPN, resulting in a constellation robust to $F_n$, $P_{in}$ and laser LW uncertainties. For each training batch, the noise figure values are sampled from a continuous uniform distribution $F_n \in [5,7]$~dB, the launch power is sampled from a continuous uniform distribution  $P_{in} \in [-2, 2]$~dBm and the RPN variance is sampled from a log-uniform distribution $\sigma^2_{RPN} \in [0.005, 0.05]$.
\end{description}

\section{Numerical results} \label{results}

The size of the constellation is $M=64$ and the training is performed using the Adam optimizer \cite{kingma2014adam} as the backpropagation algorithm. In each training epoch, a new sample set of size $N=256 \cdot M$ is generated with uniformly distributed one-hot encoded vectors and divided into batches of size $B=32 \cdot M$. The testing was done by running $100$ simulations with $10^5$ symbols per simulation in each case. A square quadrature amplitude modulation (QAM) is used as the benchmark in this study.
The BPS algorithm parameters, number of test phases $N_s=60$ and window size $W=64$, are fixed. These parameter values were chosen so that the non-shaped QAM constellation performs well on average across the studied channel conditions. In optical communication, an external cavity laser with LW of up to 100 kHz is normally used \cite{zhou2016enabling}. The combined LW of the transmitter and receiver laser can amount up to 200 kHz. In this paper, the considered region is $\Delta \nu \in [50, 250]$~kHz in order to allow a slight margin in the high-end of supported LWs. In the case of the AWGN (NLIN) model, the studied SNR interval (noise figure $F_n$ and launch power $P_{in}$ intervals) in the testing stage is (are) the same as in the training stage. 

\subsection{Additive white Gaussian noise model}

The testing results of the learned constellations from the scenarios described in subsection \ref{ss:AWGN} are denoted as AWGN~1), AWGN~2) and AWGN~3), respectively. It should be emphasized that the results for AWGN~1) are not from a single constellation but by using the constellation which obtained the best MI for the given SNR and laser LW pair. This can be viewed as the MI performance that would be achieved if we had perfect knowledge about the channel conditions at all time and the transmitter was allowed to select the constellation accordingly.

The simulation results for QAM, AWGN~1) and AWGN~2) are shown in Fig. \ref{fig:LW_robust_MI} which illustrates the MI performance with respect to laser LW for a fixed SNR~$=17$~dB. In this case, the studied laser LWs are extended to $\Delta \nu =300$~kHz to better illustrate the benefits of AWGN~2). As it was mentioned, the testing results for AWGN~1) consider the results of multiple constellations and the individual testing results for each of these constellations are shown as well. The constellations learned with a fixed variance are only beneficial for a limited range of laser LW. For example, the constellation learned with a fixed $\sigma^2_{RPN}=0.005$ could potentially be optimal for a fixed LW $\Delta \nu = 100$~kHz. However, this constellation is sub-optimal for larger laser LWs. The AWGN~2) constellation has a slight penalty compared to AWGN~1) and it is not optimal for any of the observed LWs. However, it achieves gain compared to QAM over the whole observed LW interval, with a maximum gain of 0.15 bits/symbol. From Fig. \ref{fig:LW_robust_MI}, it can be seen that a certain RPN variance is near-optimal for a range of LWs, which is due to the chosen step. However, if the LW sweep was with a step of $100$~kHz instead of $50$~kHz there would be a unique relation: RPN variances 0.005, 0.01 and 0.02 would have been optimal for the chosen LWs of $100$, $200$ and $300$~kHz, respectively.

In Fig. \ref{fig:LW_SNR_robust_MI}, the MI performance of QAM, AWGN~1) and AWGN~3) as a function of SNR for different laser LW values is shown. Only the results for laser LWs $\Delta \nu \in \{50, 150, 250\}$~kHz are shown to avoid showing similar results. The performance of a constellation labelled "non-robust" that is optimized for the pair SNR~$=18$~dB and $\Delta \nu =150$~kHz is shown as well. Additionally, iterative polar modulation (IPM) \cite{Djordjevic2010} is included in the comparison as a near-optimal constellation shape for the AWGN channel. The constellation AWGN~3) has a similar trend for all of the observed laser LW values. For low SNR values it achieves gains comparable to the constellation with perfect knowledge of the channel conditions. The gain reduces as the SNR increases but there is never a penalty compared to QAM. Up to 0.3 bits/symbol gain is achieved by AWGN~3) compared to QAM. The constellation AWGN~3) is robust to both SNR and laser LW uncertainties. The IPM constellation shows comparable performance to the optimized constellations for high SNR values and low laser LW. However, with the degradation of SNR or laser LW the performance of IPM significantly deteriorates.

The non-robust constellation has superior performance compared to AWGN~3) and almost no penalty compared to AWGN~1) for SNR~$\geq 17$~dB at laser LWs $\Delta \nu =50$ and $\Delta \nu =150$~kHz. By only observing narrow intervals of SNR~$\in [17,20]$~dB and laser LW $\Delta \nu \in [50,150]$~kHz, it can be noticed that the non-robust constellation is almost optimal. This result implies that for narrower intervals, a single constellation with close to optimal performance for given channel variations could be learned. For these laser LWs, the penalty increases with the degradation of SNR. Observing the results for laser LW $\Delta \nu =250$~kHz, the non-robust constellation does not even outperform QAM in the studied SNR region. This constellation demonstrates good performance for some of the cases, however it does not exhibit robustness to the desired degree.

\subsection{Nonlinear interference noise model}

Similar to the previous channel model, the testing results of the learned constellations from the scenarios described in subsection \ref{ss:NLIN} are denoted as NLIN~1), NLIN~2) and NLIN~3), respectively. It should be emphasized that the results for NLIN~1) are not from a single constellation but by using the constellation which obtained the best MI for the given $F_n$, $P_{in}$ and laser LW triplet. It should be viewed the same way as AWGN~1). Also, each of the constellations from NLIN~2) have been tested only on the noise figure they have been trained on. A dual polarization transmission, where both polarizations have the same constellation, was modeled but the MI performance per polarization is reported. 

\begin{figure*}[!t]
\centering
\includegraphics[width=\linewidth]{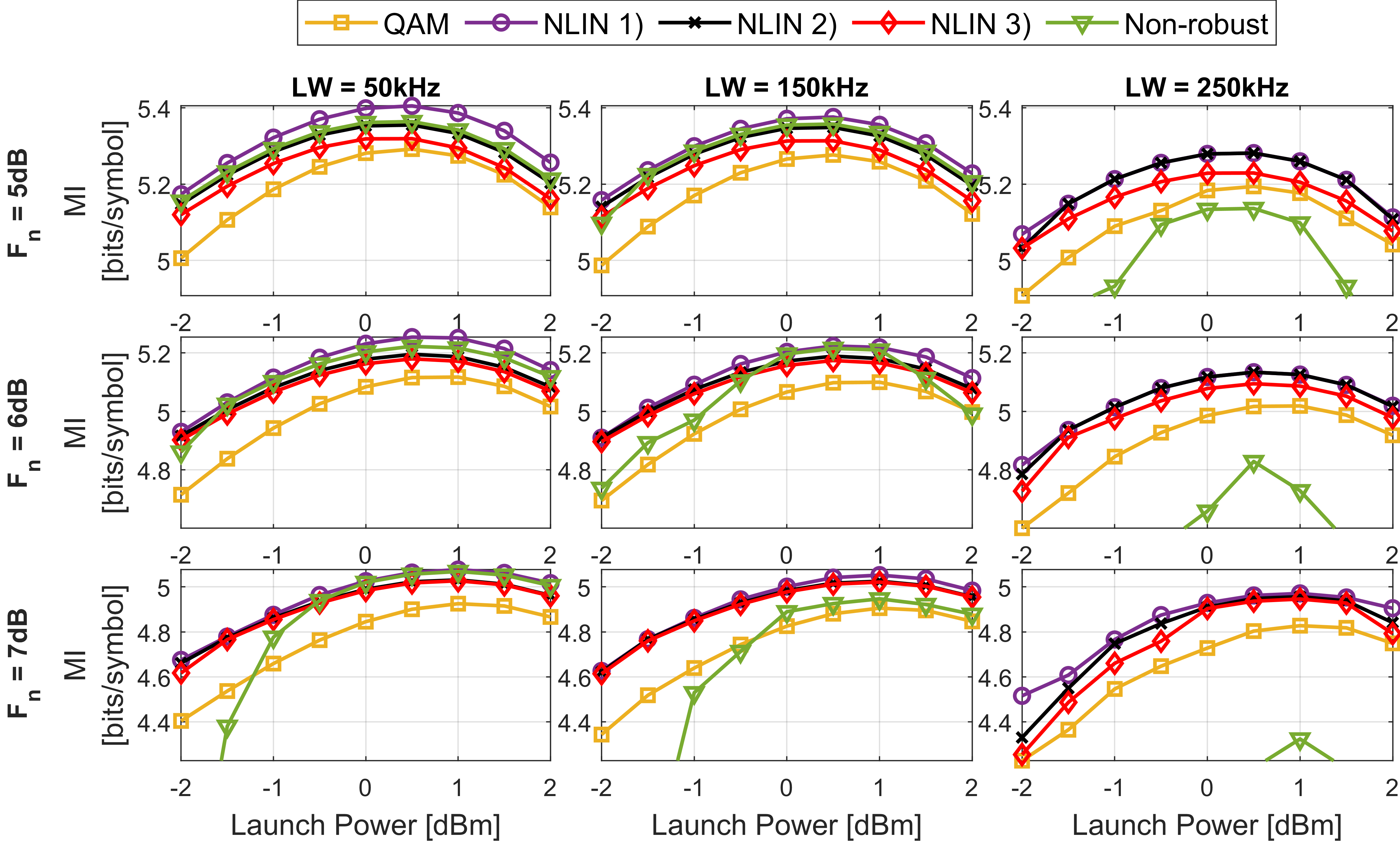}
\caption{Performance in MI with respect to launch power for different LW (columns) and NF (rows) values for QAM, NLIN~1), NLIN~2) and NLIN~3), where NLIN~1), NLIN~2) and NLIN~3) represent scenarios 1), 2) and 3) for the NLIN model, respectively. The NLIN 1) is the scenario where $P_{in}$, $F_n$ and the RPN are fixed values during training. The NLIN 2) is the scenario where $F_n$ is fixed, and $P_{in}$ and the RPN values vary during training. The NLIN 3) is the scenario where all three parameters,$P_{in}$, $F_n$ and the RPN, vary during training. The constellation denoted as non-robust is optimized for the triplet $P_{in}=0$~dBm, $F_n=6$~dB and $\Delta \nu =150$~kHz.}
\label{fig:NLIN_robust_MI}
\end{figure*}

In Fig. \ref{fig:NLIN_robust_MI}, the MI performance of QAM, NLIN~1), NLIN~2) and NLIN~3) as a function of the launch power $P_{in}$ for different noise figures (rows) and laser LW values (columns) is shown. As in the case of the AWGN model, only the results for laser LWs $\Delta \nu \in \{50, 150, 250\}$~kHz are illustrated. The performance of a constellation labelled "non-robust" that is optimized for the triplet $P_{in}=0$~dBm, $F_n=6$~dB and $\Delta \nu =150$~kHz is shown as well.
The NLIN~2) achieves substantial gain compared to QAM and has a slight penalty compared to NLIN~1) in most of the cases. Observing the results for laser LWs $\Delta \nu =50$ and $\Delta \nu =150$~kHz, the gains NLIN~2) achieves for lower studied launch powers are comparable to NLIN~1). Therefore, the NLIN~2) is exhibiting similar behavior as AWGN~3) since a lower launch power is equivalent to a lower SNR value. Furthermore, the same can be noticed when increasing the noise figure which is also equivalent to reducing the SNR. Therefore, as the launch power decreases and the noise figure increases the penalty compared to NLIN~1) becomes marginal for laser LWs $\Delta \nu =50$ and $\Delta \nu =150$~kHz. However, the same behavior cannot be seen when observing $\Delta \nu =250$~kHz. The results for NLIN~2) overlap with NLIN~1) with the exception of the lower studied launch powers for which there is a penalty. Overall, the results for NLIN~2) indicate that the learned constellations for each of the noise figures have shaping gains compared to QAM and that they are robust to launch power and laser LW uncertainty.

\begin{figure}[!t]
\centering
\includegraphics[width=\columnwidth]{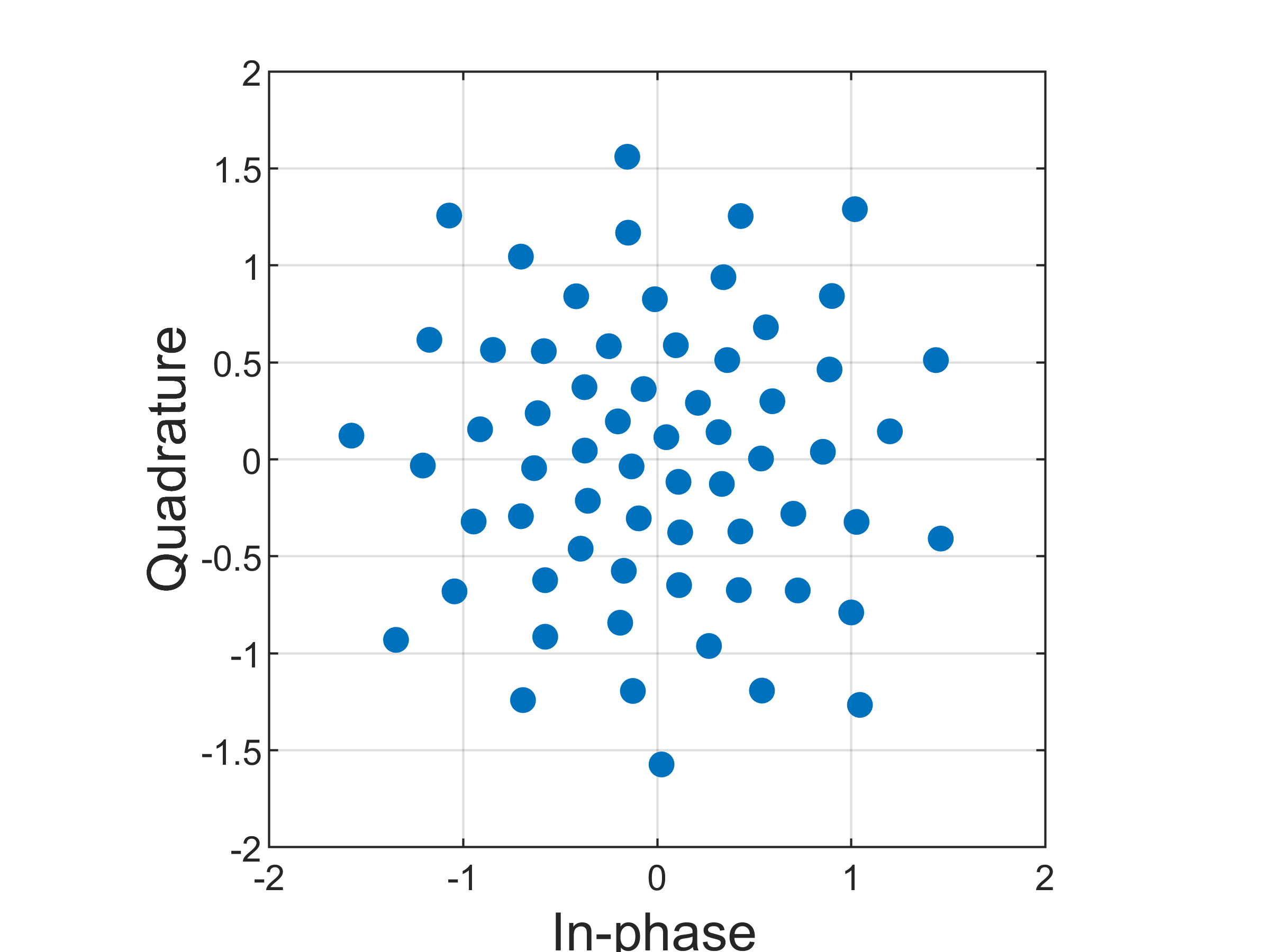}
\caption{Constellation NLIN 3) which is robust to launch power, noise figure and laser LW uncertainties.}
\label{fig:NLIN_constellations}
\end{figure}

The constellation NLIN~3), shown on Fig. \ref{fig:NLIN_constellations}, is trained to be robust to launch power, noise figure and laser LW variations. In all of the observed test cases the MI performance of constellation NLIN~3) is superior to regular QAM. The highest gain achieved by NLIN~3) compared to QAM is $0.27$ bits/symbol. For lower SNR values the constellation NLIN~3) achieves the highest gains, therefore it exhibits similar behavior to NLIN~2) and AWGN~3). Based on the results for NLIN~3), it can be concluded that the learned constellation is robust to variations in all three observed parameters, launch power, noise figure and laser LW. When comparing NLIN~3) to NLIN~2) for $F_n=6$ and $F_n=7$~dB, the differences in MI performance are marginal with the exception for $\Delta \nu =250$~kHz where NLIN~3) has penalty compared to NLIN~2) for some launch powers. In the case of  $F_n=5$~dB, the NLIN~3) is clearly inferior to NLIN~2).

Finally, let us observe a non-robust constellation that is close to optimal for the triplet $P_{in}=0$~dBm, $F_n=6$~dB and $\Delta \nu =150$~kHz. For laser LW $\Delta \nu =50$~kHz with noise figure $F_n=5$ and $F_n=6$~dB, the non-robust constellation has greater performance than NLIN~3). This can be also seen for $\Delta \nu =150$~kHz and $F_n=5$~dB, however for $\Delta \nu =150$~kHz and $F_n=6$~dB the non-robust constellation outperforms NLIN~3) only in the launch power range $P_{in} \in [0,1]$~dBm. Similarly to the AWGN model, by only observing narrow intervals of launch power $P_{in} \in [0,1]$~dBm, noise figure $F_{n} \in [5,6]$~dB and laser LW $\Delta \nu \in [50,150]$~kHz, it can be noticed that the non-robust constellation achieves performance close to NLIN~1) which is almost optimal. In the rest of the cases, the non-robust constellation does not even outperform the QAM for all of the observed launch power. Observing the results for laser LW $\Delta \nu =250$~kHz, the non-robust constellation has significant penalty to QAM.

\begin{figure}[!t]
\centering
\includegraphics[width=\columnwidth]{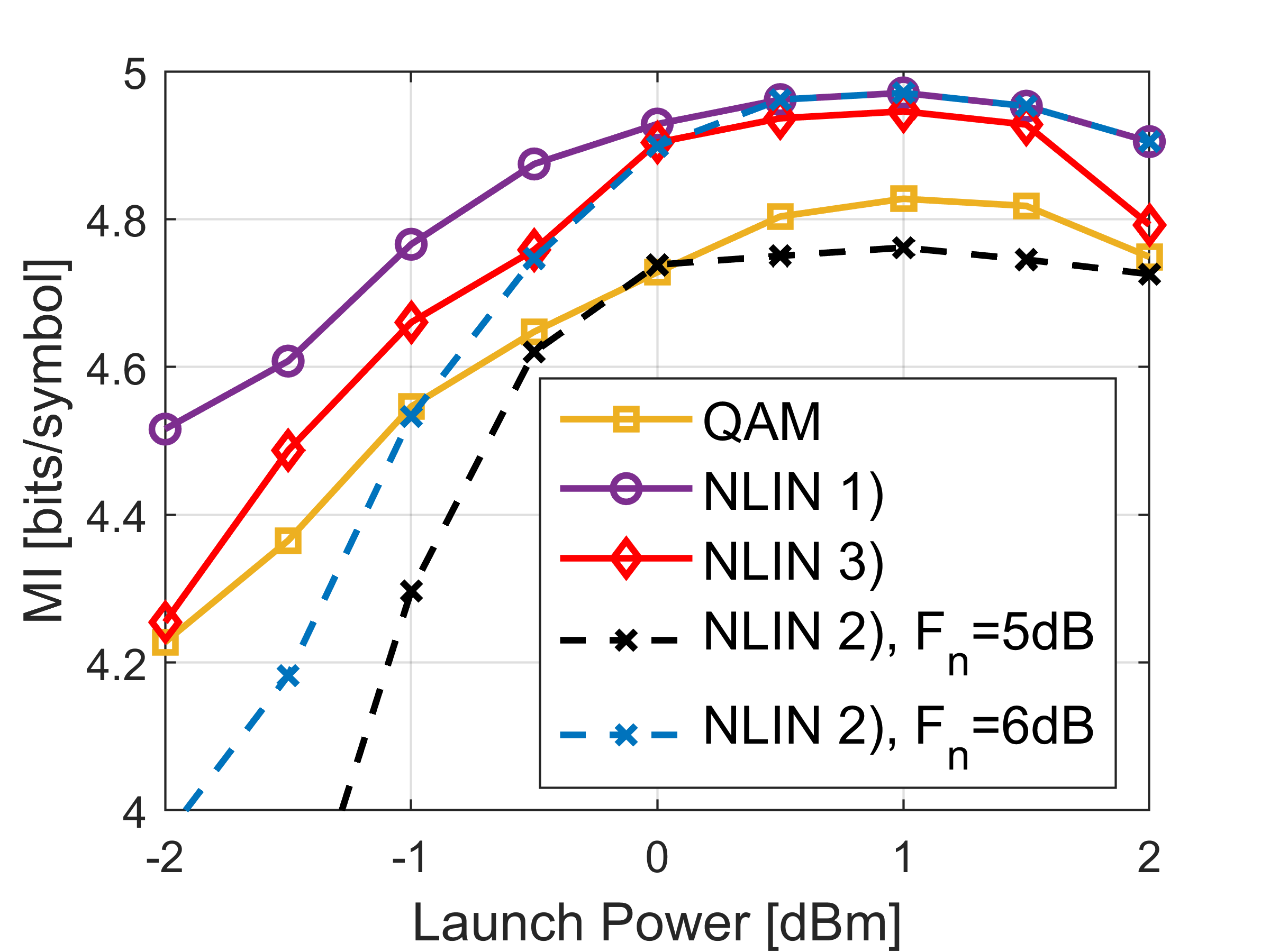}
\caption{Performance in MI with respect to launch power for $F_n=7$~dB and $\Delta \nu =250$~kHz for QAM, NLIN~1), NLIN~3), NLIN~2) trained on $F_n=5$~dB and NLIN~2) trained on $F_n=6$~dB.}
\label{fig:NLIN_7dB}
\end{figure}

For each of the noise figures, the NLIN~2) has superior performance to NLIN~3), however this comes at the expense of lower robustness, as illustrated in Fig \ref{fig:NLIN_7dB}. In Fig \ref{fig:NLIN_7dB}, the MI performance of the NLIN~2) constellations, trained on $F_n=5$ and $F_n=6$~dB (dashed lines), as a function of the launch power $P_{in}$ for $F_n=7$~dB and $\Delta \nu =250$~kHz is shown. The constellation trained on $F_n=5$~dB has a significant penalty compared to NLIN~3) and does not even outperform QAM. The constellation trained on $F_n=6$~dB overlaps with NLIN~1) for high launch powers. However as the launch power decreases below $P_{in}=0$~dBm the MI performance of this constellation rapidly decays and it has a penalty compared to NLIN~3). The NLIN~3) is more robust than NLIN~2), showing that there is a trade--off between performance and robustness. 

These results imply that if the parameters in the optical network have small variations it would be enough to perform training on a fixed set of parameters. However, for wider variation intervals, this is not the case and robust optimization such as NLIN~2) and NLIN~3) needs to be performed. Achieving robustness to the desired degree comes at the expense of MI performance, therefore it can be concluded that there is a trade--off between performance and robustness.

\section{Conclusion} \label{Conclusion}
An autoencoder was proposed for optimization of a geometric shape that is robust to uncertainties in channel conditions, such as signal-to-noise ratio (SNR) and laser linewidth. By utilizing a simpler channel model and by imposing channel variability during the training phase, the AE can be trained to produce a constellation which is robust to the uncertainty in the channel and equipment parameters, such as amplifier noise figure, launch power and laser linewidth. The test results, obtained on a more realistic channel model have indicated that using the proposed method, a robust constellation can be learned. Two additive noise models were considered, additive white Gaussian noise (AWGN) and nonlinear interference noise (NLIN). For both noise models, the learned robust constellations achieve superior mutual information (MI) performance compared to quadrature amplitude modulation (QAM) over the studied parameter intervals. Up to 0.3 and 0.27 bits/symbol of gain with respect to QAM was achieved for AWGN and NLIN models, respectively. It can be also concluded that there is a trade--off between robustness and MI performance, meaning that robustness is achieved at the expense of MI.

\section*{Acknowledgment}

This work was financially supported by the European Research Council through the ERC-CoG FRECOM project (grant agreement no. 771878), the Villum Young Investigator OPTIC-AI project (grant no. 29334), and DNRF SPOC, DNRF123.

\bibliographystyle{IEEEtran}
\bibliography{references.bib}
\end{document}